\documentclass[journal,twoside,web]{IEEEtran}
\pdfoutput=1
\usepackage{generic}
\usepackage{amsmath,amssymb,amsfonts}
\usepackage{algorithmic}
\usepackage{graphicx}
\usepackage{textcomp}
\def\BibTeX{{\rm B\kern-.05em{\sc i\kern-.025em b}\kern-.08em
    T\kern-.1667em\lower.7ex\hbox{E}\kern-.125emX}}
\usepackage[utf8]{inputenc}

\usepackage[style=ieee,backend=biber]{biblatex}

  \graphicspath{{figures/}}
\bibliography{bibliography}

\begin{document}
\title{Light-weight sleep monitoring: electrode distance matters more than placement for automatic scoring}
\author{Kaare B. Mikkelsen$^{*1}$, Huy Phan$^2$
, Mike L. Rank$^3$, Martin C. Hemmsen$^3$, Maarten de Vos$^{4,5}$, Preben Kidmose$^1$
\thanks{1: Department of Engineering, Aarhus University. 2: School of Electronic Engineering and Computer Science, Queen Mary University of London. 3: T$\&$W Engineering, Lynge, Denmark. 4: Faculty of Engineering Science, KU Leuven. 5: Faculty of Medicine, KU Leuven.}
\thanks{* mikkelsen.kaare@eng.au.dk}
}

\maketitle
\begin{abstract}
Modern sleep monitoring development is shifting towards the use of unobtrusive sensors combined with algorithms for automatic sleep scoring. Many different combinations of wet and dry electrodes, ear-centered, forehead-mounted or headband-inspired designs have been proposed, alongside an ever growing variety of machine learning algorithms for automatic sleep scoring. In this paper, we compare 13 different, realistic sensor setups derived from the same data set and analysed with the same pipeline. We find that all setups which include both a lateral and an EOG derivation show similar, state-of-the-art performance, with average Cohen's kappa values of at least 0.80. This indicates that electrode distance, rather than position, is important for accurate sleep scoring. Finally, based on the results presented, we argue that with the current competitive performance of automated staging approaches, there is an urgent need for establishing an improved benchmark beyond current single human rater scoring.
\end{abstract}






\section{Introduction}

During an 80 year lifespan, a human spends roughly 27 years asleep. As such, it should not be surprising that sleep has a large impact on virtually every major disease category, from cardiovascular disease over psychiatric disorders to cancer \cite{perez-pozuelo_future_2020}. However, diagnosis of sleep disorders is still largely confined to dedicated sleep laboratories. 
Laboratory-based polysomnography (PSG) is  the main method to gather insight in a patient’s sleep, certainly when neurophysiological data is needed. Although sleep is an essential part of several disorders, such as neuropsychiatric disorders, the practical limitations for wide scale use of PSG hamper the integration of sleep as a vital component in diagnostic and therapeutic trajectories of patients with these disorders. Moreover, the sleep laboratory is a very artificial environment, which has an influence on sleep itself. In order to better understand the impact of healthy and abnormal sleep-wake patterns on various disease conditions, there is an urgent need for sleep monitoring over prolonged periods of time outside traditional sleep clinics.   

In the past decade, multiple studies have explored the use of digital wearable (e.g. actigraphy) and bed-side (e.g. radar-based) sensors to quantify various aspects of sleep, but failed to capture the neurophysiological signatures that underpin the quantification of sleep based on AASM convention \cite{berry_aasm_2017}. With the introduction of various wearable EEG sensors, capturing brain activity from unconventional places (e.g. behind or in the ear), personalised long-term sleep monitoring on the general population is within reach \cite{mikkelsen_accurate_2019,levendowski_accuracy_2017,
lucey_comparison_2016,looney_wearable_2016,
arnal_dreem_2020,mikkelsen_automatic_2017,
popovic_automatic_2014,mikkelsen_machine-learning-derived_2018}.

Visually reviewing the large amount of time series sleep data that could be recorded with this new generation of wearable EEG would be time-consuming and costly, in addition to requiring re-training of the human scorers for each new wearable (which would be highly inefficient \cite{mikkelsen_machine-learning-derived_2018}).  For decades, machine learning scientists have attempted to mimic visual annotation based on AASM rules by handcrafting features and training machine learning algorithms. This approach achieved only moderate success. More recently, the field of automated sleep staging embraced artificial intelligence (AI) technology, in particular deep learning (DL) architectures. This lead to a variety of promising automated sleep analysis approaches. An important advantage of automated scoring approaches is the absence of intra-scorer variability \cite{berthomier_exploring_2020}. Those automated staging approaches are primarily developed and validated on large, publicly available PSG datasets.  With minor modifications, such DL architectures could also be used for automatic staging of wearable sleep data \cite{phan_deep_2019}. However, due to the variety of available wearable sensors and the experimental nature of data collection with those devices,  'wearable' datasets are still an order of magnitude smaller compared to PSG data sets, and performance of automated staging approaches requires further investigation.

In this paper, we apply one of the leading analysis pipelines, the SeqSleepNet \cite{phan_seqsleepnet_2019}, to multiple different, realistic sensor configurations. 'Realistic' in this sense means that we only test montages where a limited number of electrode positions are used at a time, and only positions that are reasonably hidden and easy to access (out of the hair line and only on the sides of the head).

The SeqSleepNet pipeline was validated by an independent group on two different datasets, showing that the performance of the network outperformed the average human annotator \cite{guillot_dreem_2020}. In this paper, we show that this network, originally developed and trained for automatically staging PSG, can be directly applied to in-ear EEG data.  In addition, we investigate the likely upper limits to mobile sleep scoring accuracy and the variations between different	 approaches.

\section{Methods}
\subsection{Data}
\begin{figure}
 \includegraphics[width=\linewidth]{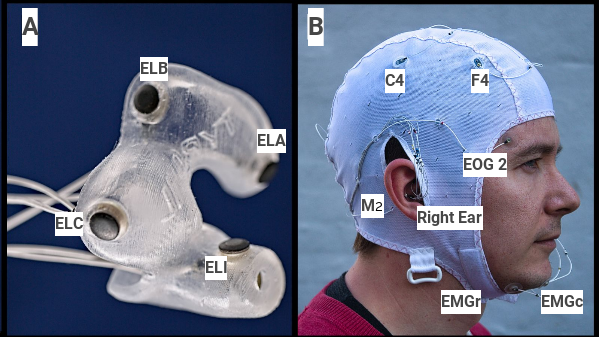}
 \caption{The recording setup used in this study. A: example of the soft ear-EEG electrode holders, with embedded dry-contact electrodes, placed in each ear. B: cap-mounted PSG setup using 8 scalp EEG electrodes, 2 EOG electrodes and 3 EMG electrodes. See \cite{mikkelsen_accurate_2019} for a detailed description.}
 \label{fig:setup}
 \end{figure}

 We used the 80 nights of sleep recordings (4 nights from 20 subjects each) which were presented in Mikkelsen et al 2019\cite{mikkelsen_accurate_2019}. This data set consists of concurrent PSG (13 electrodes) and ear-EEG (6 electrodes in each ear) recordings. See Figure \ref{fig:setup} for an example of the setup.
 
 Rather than using the raw data, we work with the sleep recordings after artefact rejection, as described in Mikkelsen et al \cite{mikkelsen_accurate_2019}. In this artefact rejection pipeline, artefacts are identified on an individual electrode basis, and are removed by changing the relevant sample values to 'NaN' (which enables discarding samples from individual channels). During preparation of the various derivations, NaN-values are ignored when EEG electrodes are averaged (as is the case with ear derivations). If there were any NaN's in a final derivation, the missing samples were linearly interpolated from the nearest non-missing values. For extended missing sections, the interpolated values decayed exponentially towards zero (with time scale 1 second).
 
 The PSG recordings have been scored by two independent and experienced sleep technicians ('scorer 1' and 'scorer 2'), according to the AASM guidelines \cite{berry_aasm_2017}. We have decided to treat scorer 1 as the ground truth, to which the automatic sleep classifiers will be compared (and trained on). In contrast, scorer 2 is an independent source of labels, which will be used in studying the possible causes of classifier errors. 
 
\subsection{Choice of electrode configurations and epochs}

\begin{figure*}[htbp]
\centering
\includegraphics[width=.8\textwidth]{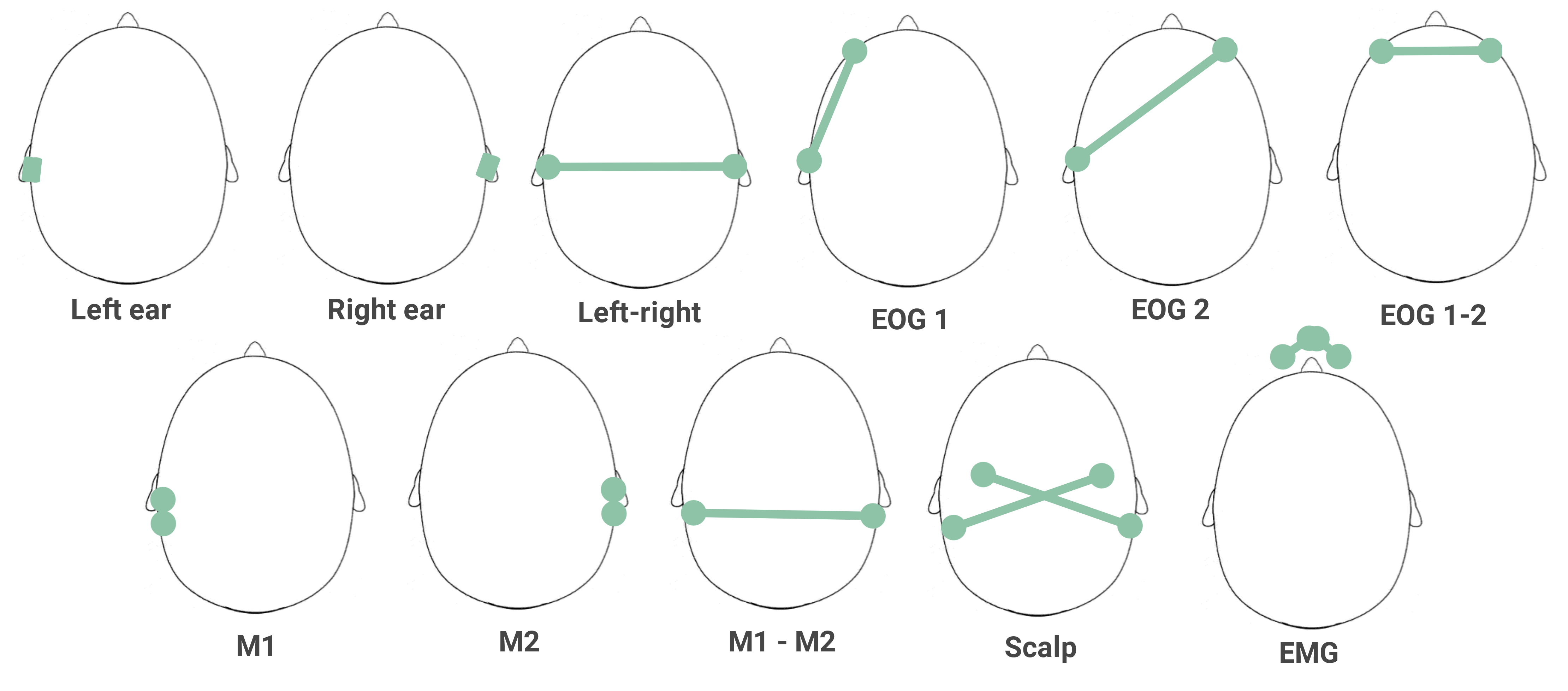}
\caption{Overview of different derivations used. 'Left ear' and 'right ear' uses the average over the three innermost ear electrodes versus the average of the three outermost electrodes, in each ear. Left-right use the average of all electrodes in each ear. 'M1', 'EOG 1' and 'EOG 2' references a single electrode to the average of all left ear electrodes. For 'Scalp', both C3-M2 and C4-M1 are calculated; for each recording we used the derivation with the least rejected or lost samples. For 'Chin EMG', all three derivations between all three EMG electrodes (l, r, c) are calculated. If l-r has a missing sample, r-c is used instead. If r-c is missing as well, l-c is used. }
\label{fig:eegDeriv}
\end{figure*}

Figure \ref{fig:eegDeriv} shows all electrode derivations under consideration in this study. As can be seen, we have chosen to rely more on the left than right side of the head. This was done both to reduce the number of derivations at play, and because previous work had shown the left ear electrodes were found to be slightly more reliable than the right ear electrodes  \cite{mikkelsen_accurate_2019}, we will elaborate more on this in the 'Results' section. In designing the 'Scalp' and 'EMG' derivations, we decided to try to make them as reliable as possible, by combining multiple derivations in one. This was done because we are primarily interested in the performance of a mobile sleep monitoring setup, and we do not consider chin EMG or scalp EEG electrodes to be prime candidates for user friendly mobile setups. Therefore, the primary concern for these data channels is that they are responsible for as little data rejection as possible.

\begin{figure*}[htbp]
\centering
\includegraphics[width=.7\textwidth]{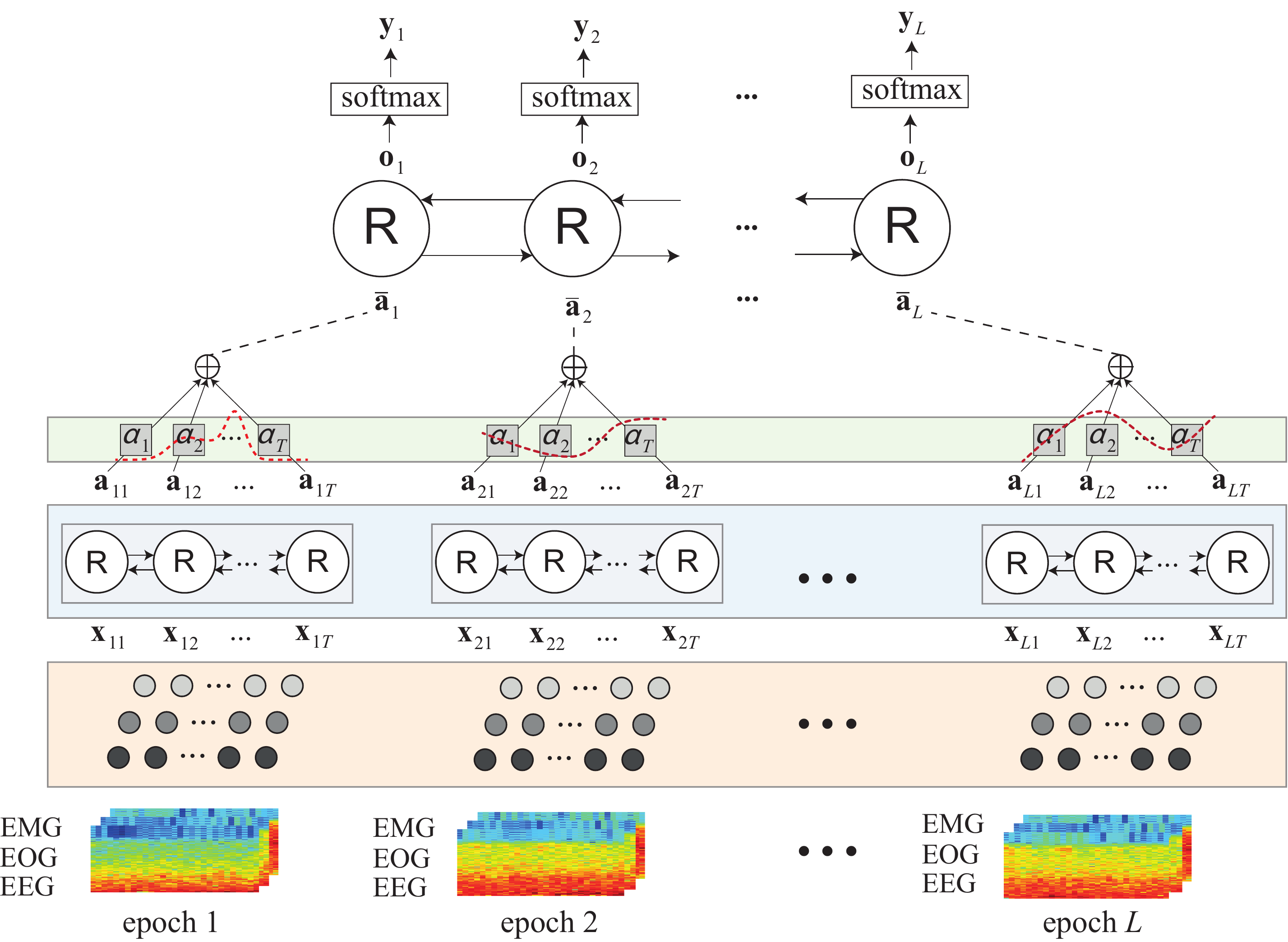}
\caption{Illustration of SeqSleepNet. The figure is adapted from \cite{phan_seqsleepnet_2019}. In this paper, $L=20$.}
\label{fig:seqsleepnet}
\end{figure*}

\subsubsection*{Epoch rejection}

 To make the comparison of different setups (meaning different combinations of derivations) as unambiguous as possible, \textbf{ we only use epochs for which all derivations are well defined}. In this regard, a derivation is considered 'ill defined' if all samples in that epoch for that derivation have been rejected (replaced with 'NaN' values). In cases when a derivation is constructed by averaging a set of channels, any 'NaN'-values of an individual channel are ignored. 
 
 Using these statistics, we evaluate whether any derivations should be excluded from the analysis. In this regard, the important metric is not the individual reliability of the derivation, but rather to which degree the derivation is well-defined when other derivations are. If it is not, it will be directly responsible for reducing the number of viable epochs. As is shown later, we end up removing the 'right ear' derivation.

\subsection{The SeqSleepNet classifier}

In this study we used SeqSleepNet \cite{phan_seqsleepnet_2019}, illustrated in Fig. \ref{fig:seqsleepnet}, as the base classifier. SeqSleepNet works by analysing a sequence of $L$ consecutive epochs and classifying them at once into a sequence of $L$ sleep stage labels (i.e., sequence-to-sequence). We set $L = 20$ in this study as recommended in \cite{phan_seqsleepnet_2019}. The data input to the network can be single- or multiple- channel log-scale spectrograms. The data of each channel was normalized to have zero mean and unit variance for each frequency bin using the normalization parameters computed from the training data.

The $i$-th epoch, $1 \le i \le L$, in the input sequence was encoded into a feature vector $\bar{\mathbf{a}}_i$ via the epoch encoder. The epoch encoder is composed of (1) filter-bank layers, one for each input channel, (2) a bidirectional recurrent layer realized by a long short-term memory (LSTM) cell, and (3) an attention layer. The spectrogram channels first have their frequency dimension smoothed and reduced via the filter-bank layers. The filtered spectrograms are then stacked along the frequency direction and presented to the LSTM, which converts them into a sequence of output vectors. The output vectors in this sequence are eventually combined, using weights learned by the attention layer to form the feature vector $\bar{\mathbf{a}}_i$.  

Going through the epoch encoder, the input sequence was transformed into a sequence of feature vectors. An LSTM-based bidirectional recurrent layer was then employed for inter-epoch sequential modelling, converting the sequence of feature vectors into a sequence of output vectors. These output vectors were finally presented to a fully-connected layer, followed by a softmax layer, for classification, producing a sequence of labels, each label corresponding to an epoch in the input sequence. The network was trained end-to-end to minimize the cross-entropy loss averaged over the sequence. See \cite{phan_seqsleepnet_2019} for more details.

\subsection{Classifier training and transfer learning}
\subsubsection*{Training}
To test a wide selection of different, relevant electrode combinations, we used different subsets of the electrode derivations as inputs to the network, these are listed in Table \ref{tab:123input}. Here, `+' means that multiple derivations are given as separate inputs. The SeqSleepNet was configured similar to original implementation \cite{phan_seqsleepnet_2019}. 

\begin{table}[]
\caption{Overview of channel combinations used. '+' means using mutiple, separate data channels.}
\label{tab:123input}
\centering
\begin{tabular}{c|c|c}
\multicolumn{3}{c}{Single, double or triple channel input:} \\
\multicolumn{1}{c}{Single} & \multicolumn{1}{c}{Double} & \multicolumn{1}{c}{Tripple}                                             \\\hline            

\begin{tabular}[c]{@{}l@{}}'Left ear'\\ 'M1'\\ 'EOG 1-2'\\ 'LR' \\ 'M1-M2' \\ 'scalp' \end{tabular} & \begin{tabular}[c]{@{}l@{}}'M1'+'EOG1'\\ 'M1'+'EOG2'\\ 'LR'+'EOG1'\\ 'LR'+'EOG2'\\ 'LR'+'EOG 1-2'\end{tabular} & \begin{tabular}[c]{@{}l@{}}'scalp'+'EOG 1-2'+'EMG'\\ 'LR'+'M1'+'M2'\end{tabular}                        
\end{tabular}
\end{table}

\subsubsection*{Transfer learning}

 As an alternative to training directly on the reduced electrode set, we also studied the effect of transfer learning \cite{phan_towards_2020}. To this end, we pretrained SeqSleepNet with the Montreal Archive of Sleep Studies (MASS) database, which consists of 200 subjects \cite{oreilly_montreal_2014}. For this test, only a single-input version of the network was prepared, using the C4-A1 derivation. The pretrained networks were then used as the starting points and further trained (i.e., the entire network were finetuned) with our data.

\subsubsection*{Performance evaluation}

In the remainder of this paper, we shall refer to a SeqSleepNet trained for a specific set of inputs as a `classifier'. When discussing both manual sleep scorers and automatic classifiers, we shall refer to to all of them as `scorers'.

Each classifier was trained and tested in a leave-one-subject-out fashion, using 15 subjects for training, 4 subjects for validation and one for testing. For each subject, all available recordings were used. 
 
 When quantifying classifier performance, we use Cohen's kappa \cite{cohen_coefficient_1960}.

\section{Results}

\begin{figure*}[htbp]
\includegraphics[width=1\linewidth]{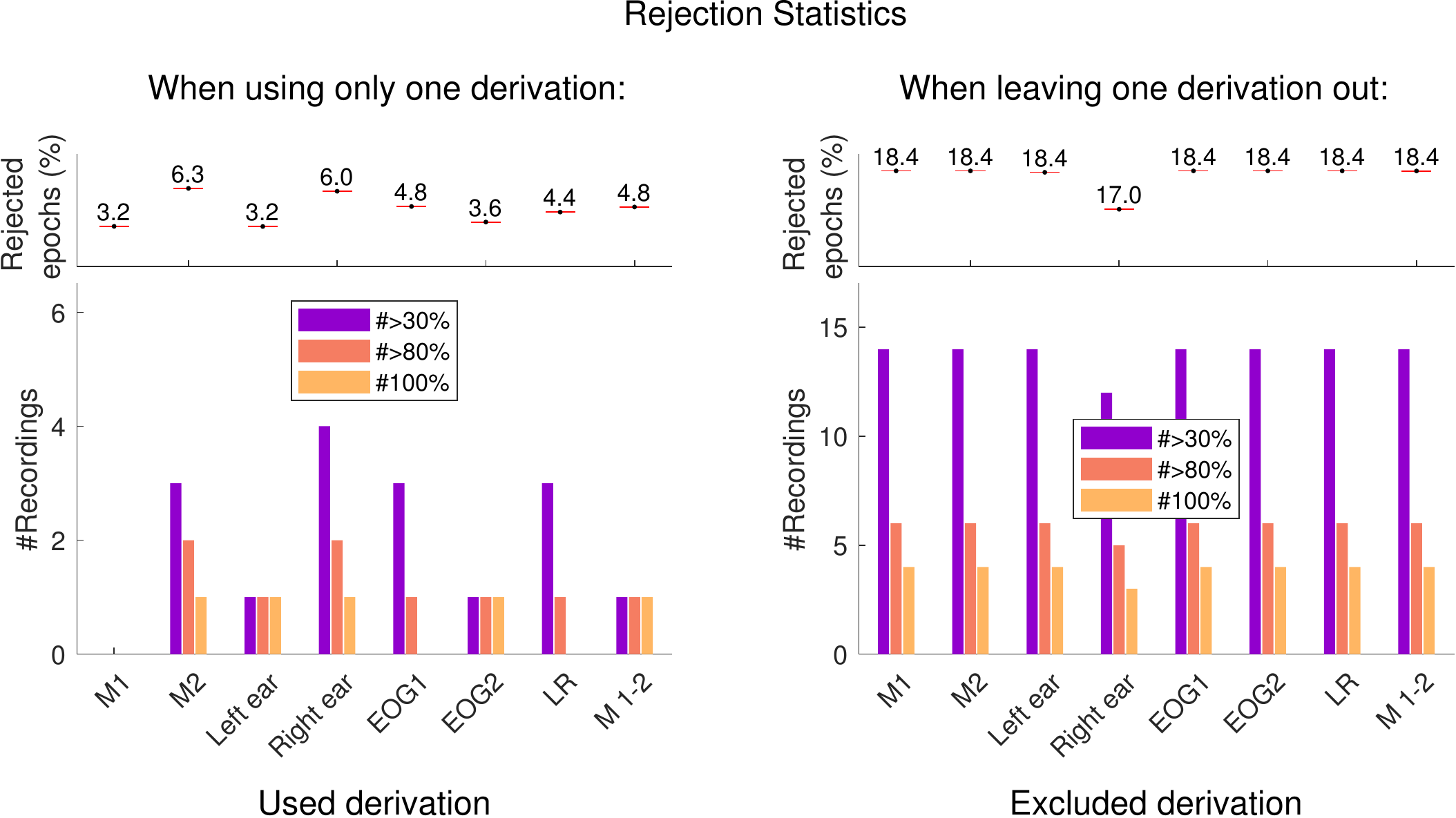}
\caption{The relationship between data channels used and amount of accepted epochs. Left, top: percentage of epochs rejected based on a single data channel. Left, bottom: number of recordings for which at least 30, 80 and 100 percent of epochs were rejected based on a single data channel. Right, top: percentage of epochs rejected when all data channels except one were included (the excluded channel is shown on the bottom x-axis). Right, bottom: number of recordings for which at least 30, 80, and 100 percent of epochs were rejected when all data channels except one were included. }
\label{fig:confQual}
\end{figure*}

Figure \ref{fig:confQual} shows how the set of accepted epochs depends on the chosen set of derivations.
On the left is shown rejection statistics for individual derivations. We see that the epoch-wise rejection rate is quite stable across derivations, varying between 3.2$\%$ and 6.0$\%$. For comparison, scorer 1 marked $3.5\%$ of the epochs as unclassified. However, more important than the single derivation statistics is the impact on epoch rejection when multiple derivations are considered. On the right, statistics are shown for when all but one derivation are used. Again, we see that excluding a single derivation mostly does not change the overall rejection rate. However, we note that removing the 'right ear' derivation reduces the number of recordings that are completely rejected (from 3 to 2). Because of this, we decided to exclude the 'right ear' derivation from the rest of the analysis, and use the $83\%$ of epochs which are accepted in all other derivations (including being scored by scorer 1). This results in excluding 2 recordings (from two different subjects).

 Please note that EMG, EOG and Scalp  derivations have been excluded from the comparison in Figure \ref{fig:confQual}. This is because these derivations are all necessary to perform our analysis (constituting a three-channel PSG classifier), and thus their inclusion is obligatory. 

\begin{figure}[htbp]
\includegraphics[width=1\linewidth]{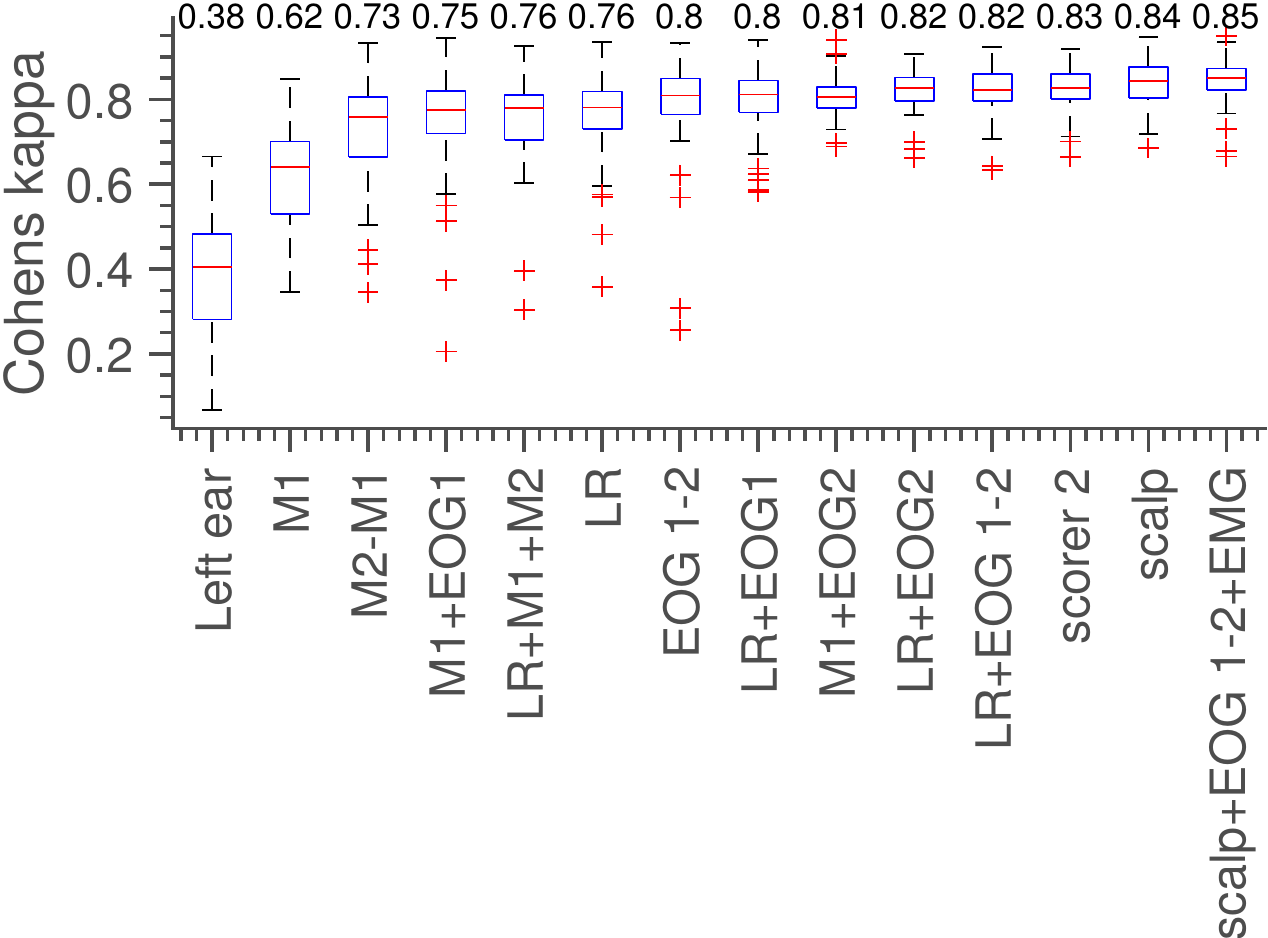}
\caption{Distributions of all kappas for all scratch-trained classifiers, relative to the labels from scorer 1. Numbers at the top are averages.}
\label{fig:kappa_scratch}
\end{figure}

Figure \ref{fig:kappa_scratch} shows boxplots for distributions of Cohen's kappa between the classifier output and the manual labels assigned by scorer 1. It is interesting to note how any classifier which combines both lateral and EOG information reaches kappa values of about 0.76 or above (in particular how well 'EOG 1-2' performs). Also, we see that the 'scalp+EOG 1-2+EMG'-classifier actually reproduces scorer 1 better than scorer 2 does. This indicates that SeqSleepNet manages to incorporate the special quirks of scorer 1, and that it is probably unwarranted to attempt further improvements in PSG-based scoring (at least when training against a single scorer).

 In the case of transfer learning, we tested the effect on the 'LR', 'LR+EOG1', 'LR+EOG2' and 'M1-M2'. We found average increases in Cohen's kappa of 0.016, 0.009, 0.005 and 0.029, respectively. 

\begin{figure*}[htbp]
\includegraphics[width=1\linewidth]{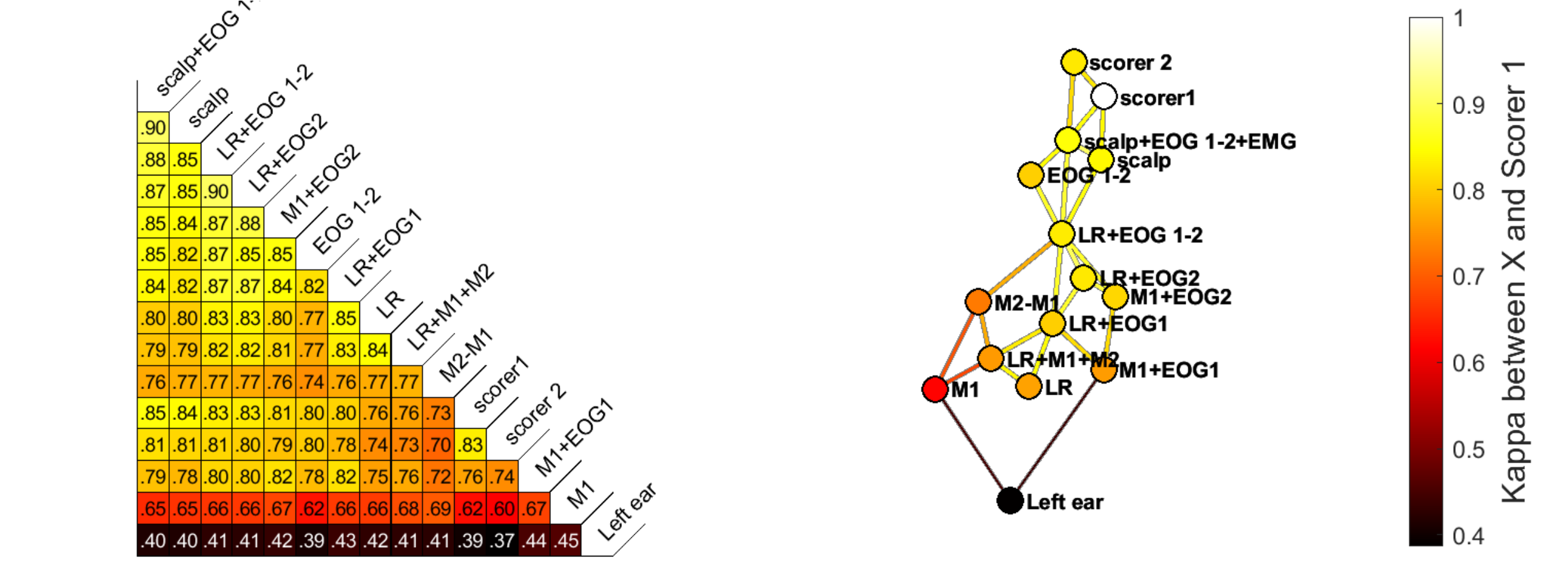}
\caption{Left: Matrix showing all pairwise Cohens kappa coefficients, ordered to maximise nearest neighbour values.  Right: Graph ordering scorers based pairwise kappas. Each node represents a scorer, and the edge weights represent the kappa value between the two scorer outputs. The node color shows the kappa value between classifier output and scorer 1 labels. For clarity, only the two strongest edges for each node have been included.}
\label{fig:graph1}
\end{figure*}

Figure \ref{fig:graph1} shows a visual comparison between all scorers, both manual and automatic. For each scorer, all kappa values are calculated relative to all other scorers (by bundling all recordings into one, and calculating one total kappa value), and the two highest values are plotted as edges on the graph. This means that while some nodes (each representing a scorer) have more than two connected edges, all nodes have at least two. The edges are coloured depending on the kappa value, and the nodes are coloured depending on the kappa value between the scorer and scorer 1.

An interesting observation can be made from Figure \ref{fig:graph1}: even though 'scorer 1' is the target that all automatic classifiers are aiming at, they do in fact agree more with each other than with scorer 1. This happens even for classifiers that do not have any input derivations in common (e.g. 'M1+EOG 2' and 'LR+EOG 1' may share electrodes, but not derivations). In particular, we note that even though 'scalp+EOG 1-2+EMG' attains the highest kappa relative to scorer 1 of any scoring method, it still attains even higher kappa values with other automatic classifiers. We can think of two plausible causes of this: (1) the manual scorer likely also makes some mistakes, meaning that there is an upper limit to how well an entirely rules-based sleep classifier can predict manual scoring. (2) it is possible that the manual sleep scorer uses information not available to the classifiers - either because the manual scorer considers more than the last 5 minutes of recording when scoring a given epoch, or because they consider other aspects of the recording, such as time of night, total duration etc., which are not revealed to the automatic classifier. 

\begin{figure}[htbp]
\includegraphics[width=1\linewidth]{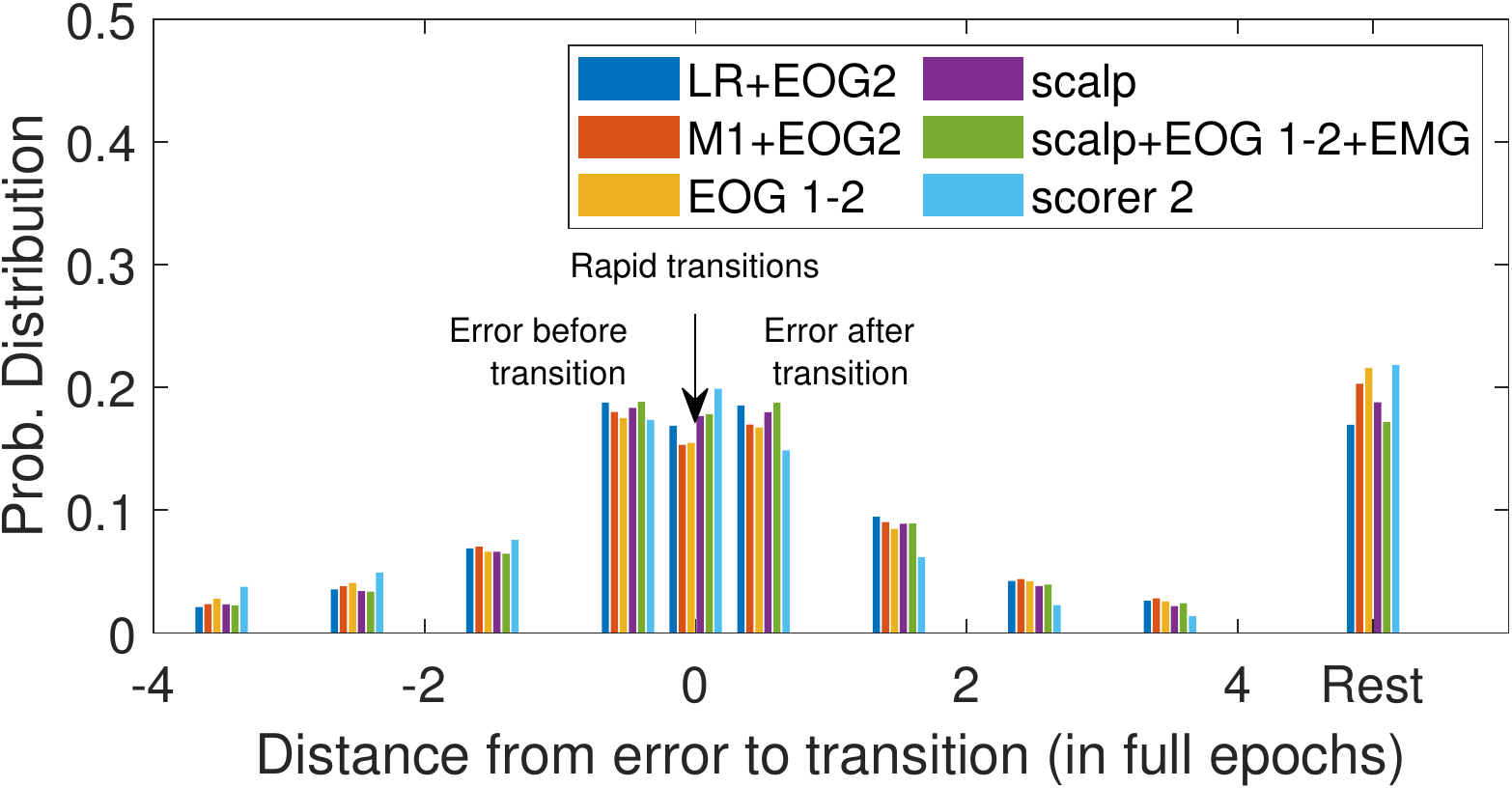}
\caption{Distribution of automatic classifier errors as a function of distance to nearest stage transition (as defined by the scoring from manual scorer 1). In the figure, 'Rapid transitions' refers to the scenario where scorer 1 only spends a single epoch in the given stage, meaning that possible error is both immediately before and after the transition.}
\label{fig:errorVsTrans}
\end{figure}

 When we further analyse the discrepancies between manual and automatic scoring, we find, not surprisingly, that most errors happen close to state transitions. This is shown in Figure \ref{fig:errorVsTrans} where we see that almost $60 \%$ of discrepancies between manual and automatic scoring happens immediately before or after a stage transition (as judged by scorer 1). Including three additional epochs to either side of the transition brings the total up to around $80\%$. Not shown is a comparison to the distance-to-transitions for a randomly chosen epoch. We find that the distributions are significantly different, which discredits the hypothesis that errors might be unrelated to transitions.

\begin{figure}
\includegraphics[width=\linewidth]{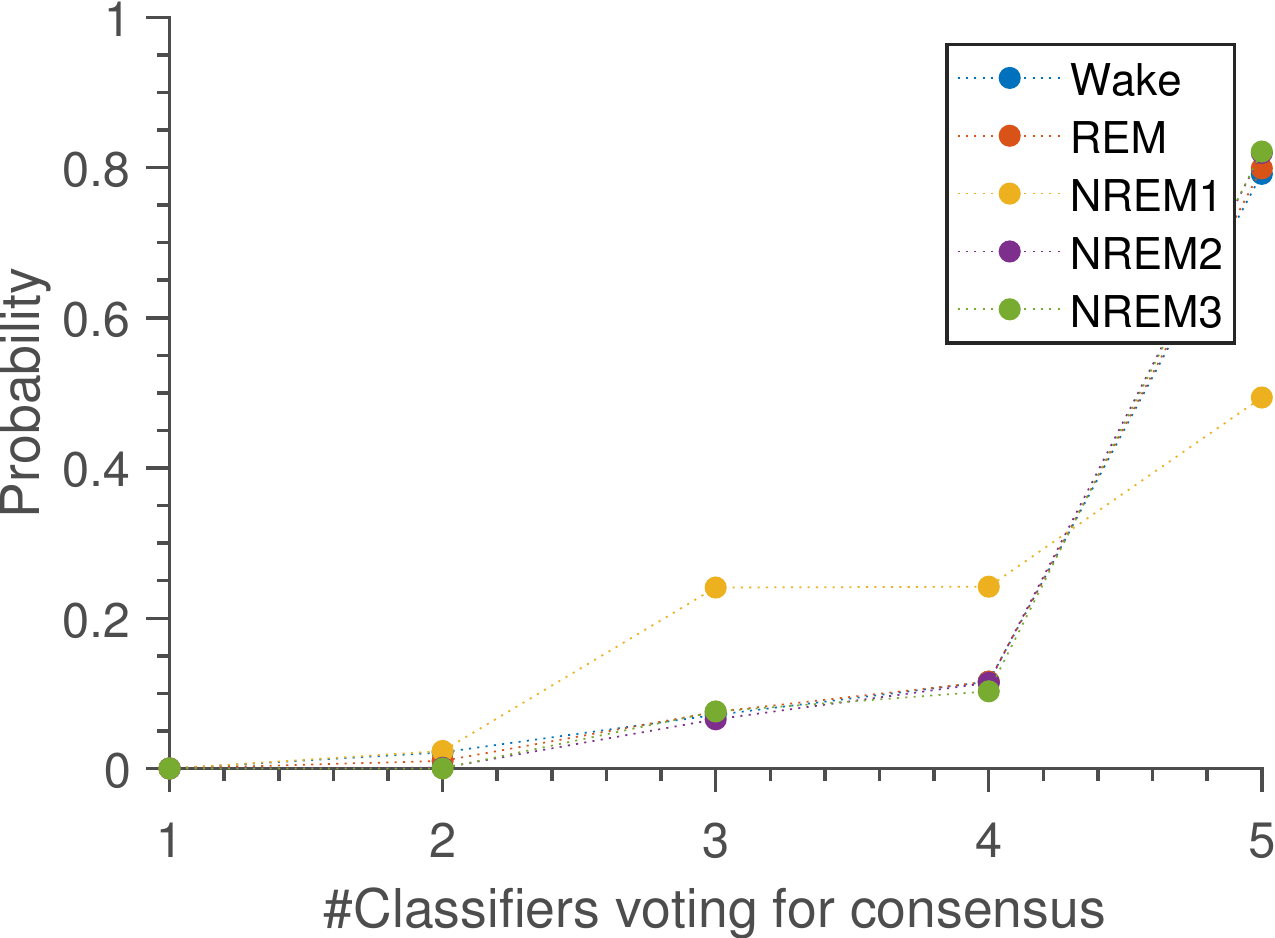}
\caption{Among the 6 best automatic classifiers, we count how many of them agree with their own consensus label, for each epoch. We see that in the overwhelming majority of epochs, all 6 classifiers agree. The one clear outlier is NREM1, which it much less well defined than the others.}
\label{fig:scorerCons}
\end{figure}

Given the high agreement between many of the automatic classifiers, we decided to specifically study the level of consensus between some of the most well-performing classifiers. We chose the following 5: 'LR + EOG2','EOG 1-2','M1 + EOG2','scalp','scalp + EOG 1-2 + EMG'. When comparing each of the 5 classifier outputs to their own majority vote, we overwhelmingly find that the 6 classifiers mutually agree. Figure \ref{fig:scorerCons} shows the average number of votes for the majority (maximum 5) for different sleep stages (as judged by the majority). We see that in $80\%$ of cases there is complete consensus, except for stage N1, which is also considered the least well-defined stage.

\section{Discussion and Conclusion}

 By combining a high number of recordings with an advanced sleep scoring algorithm, we achieve a consistent, high scoring performance, as measured by the Cohens kappa value. In fact we generally achieve a higher kappa value than what has previously  been found for mobile sleep setups \cite{mikkelsen_accurate_2019,arnal_dreem_2020}. This is central for the realization of light weight sleep monitoring, and our results here show that this can be reality. Additionally, importantly, we find that a broad selection of electrode placements,  all having in common that they have both lateral and EOG components, achieve very similar performances. This means that electrode placements should  be chosen based on unobtrusiveness, reliability and comfort, and if the recording setup is otherwise sound, we predict that a very large number of different sensor combinations can make a viable sleep monitor.

In particular, we found that a PSG-based automatic scorer, which performed very well in reproducing scorer 1, still had a higher Kappa value with at least two other automatic classifiers. This indicates that the high internal consistency among the automatic classifiers is not entirely due to limited sleep information in the non-PSG derivations, but is likely also related to the human peculiarities in the scoring by scorer 1. Apparently, the automatic classifiers all manage to define certain special cases more consistently than scorer 1, leading to the 'scalp+EOG 1-2+EMG' classifier attaining both the highest kappa value with scorer 1, while at the same time having higher agreement with other automatic classifiers. 

Based on this observation, it would be very interesting to compare the output of the classifiers presented here with output from consensus-trained PSG-based classifiers such as the one presented in Stephansen et al 2017 \cite{stephansen_use_2017}, which the authors believe could be more consistent than the gold standard manual sleep scoring.

It is worth noting that we have found no indication that the specific choice of epochs used here (as described above) is particularly easy to score, which would introduce a bias towards artificially high kappa values. We have tested the random forest based classifier presented in Mikkelsen et al. 2019 \cite{mikkelsen_accurate_2019}, on the same, reduced epoch set, (used for both testing and training), however it only attains an average kappa coefficient of 0.72 (compared to the 0.73 which was achieved using a larger set of epochs). 

In future work, it will be interesting to see how these results change when a more challenging cohort is used - it is possible that as sleep and its associated biomarkers change with age or infirmity, the optimal electrode locations will change accordingly. 

On the topic of future directions, we feel that this work highlights the need for a change in focus regarding how machine learning is used to improve clinical sleep analysis. Given the apparent high reliability of automatic sleep scoring shown in this paper and others, we believe that the goal of reproducing manual scoring  for 'regular sleep' has been largely reached. Rather than marking any kind of end to the project of updating clinical sleep analysis, we believe this marks the beginning of a new phase. To anyone following this field, it should be clear that cost-effective, long term sleep monitoring is becoming a reality. The question now is, how can automatic scoring be transformed into a trusted, clinical tool (as was recently suggested by the American Academy of Sleep Science \cite{goldstein_cathy_a_artificial_nodate}), and how can we use this tool to actually update the framework within which sleep is analyzed? For instance, the work presented in this paper would likely have benefitted from a more finegrained definition of sleep stages, such as the 'hypnodensities' that some researchers have been advocating \cite{stephansen_use_2017}. That is an example of how the existence of accurate, automatic sleep scoring, suitable for long-term monitoring, can motivate and support development of new approaches. We hope that much more of such developments are on the way.

\section{Competing Interests}

Authors MLR and MCH are employed by T $\&$ W Engineering, which develops equipment for long term EEG monitoring.

\section{Acknowledgements}
This work was sponsored by the Innovation Fund Denmark, grant 7050-00007. This research was supported by funding from the Flemish Government under the “Onderzoeksprogramma Artificiële Intelligentie (AI) Vlaanderen” programme.

\section{Author Contribution}
KBM performed the recordings and the analysis and wrote the manuscript. HP and MdV designed the sleep scoring algorithm, PK designed and build the recording equipment, KBM, MLR, MCH and PK designed the experiment, all authors participated in presenting and formulating the results.

\section{Data and code availability}
The authors are happy to share both the aggregate statistics presented in this manuscript, as well as the entire code base.

\printbibliography

\end{document}